\documentclass
[aps,pra,twocolumn,showpacs,lengthcheck,preprintnumbers]{revtex4}%
\usepackage{amsfonts}
\usepackage{amsmath}
\usepackage{amssymb}
\usepackage{graphicx}%
\providecommand{\U}[1]{\protect\rule{.1in}{.1in}}
\providecommand{\U}[1]{\protect\rule{.1in}{.1in}}
\begin{document}
\preprint{Phys. Rev. A \textbf{83}, 053610 (2011)}
\title{Gap solitons in elongated geometries: the one-dimensional Gross-Pitaevskii
equation and beyond}
\author{A. Mu\~{n}oz Mateo}
\email{ammateo@ull.es}
\author{V. Delgado}
\email{vdelgado@ull.es}
\affiliation{Departamento de F\'{\i}sica Fundamental II, Facultad de F\'{\i}sica,
Universidad de La Laguna, 38206 La Laguna, Tenerife, Spain}
\author{Boris A. Malomed}
\email{malomed@post.tau.ac.il}
\affiliation{Department of Physica Electronics, School of Electrical Engineering, Faculty
of Engineering, Tel Aviv University, Tel Aviv 69978, Israel}
\date{15 February 2011}


\begin{abstract}
We report results of a systematic analysis of matter-wave gap solitons (GSs)
in three-dimensional self-repulsive Bose-Einstein condensates (BECs)
loaded into a combination of a cigar-shaped trap and axial optical-lattice
(OL) potential. Basic cases of the strong, intermediate, and weak radial
(transverse) confinement are considered, as well as settings with shallow and
deep OL potentials. Only in the case of the shallow lattice combined with
tight radial confinement, which actually has little relevance to realistic
experimental conditions, does the usual one-dimensional (1D) cubic
Gross-Pitaevskii equation (GPE) furnish a sufficiently accurate description of
GSs. However, the effective 1D equation with the nonpolynomial nonlinearity,
derived in Ref. [Phys. Rev. A \textbf{77}, 013617 (2008)], provides for quite
an accurate approximation for the GSs in all cases, including the situation
with weak transverse confinement, when the soliton's shape includes a
considerable contribution from higher-order transverse modes, in addition to
the usual ground-state wave function of the respective harmonic oscillator.
Both fundamental GSs and their multipeak bound states are considered. The
stability is analyzed by means of systematic simulations. It is concluded that
almost all the fundamental GSs are stable, while their bound states may be
stable if the underlying OL potential is deep enough.

\end{abstract}
\maketitle



\section{INTRODUCTION}

Matter-wave gap solitons (GSs) are localized modes that can be created in
elongated Bose-Einstein condensates (BECs) loaded into one-dimensional (1D)
optical-lattice (OL) potentials, with the intrinsic nonlinearity induced by
repulsive interactions between atoms. GSs have been the topic of a large
number of original works. Results produced by these studies were summarized in
several reviews \cite{Kon1,Mor1,Carre1,Carre2,Barcelona}. Within the framework
of the mean-field approximation, the description of matter-wave patterns is
based on the Gross-Pitaevskii equation (GPE) for the macroscopic wave function
$\psi$ \cite{GPE}:%
\begin{equation}
i\hbar\frac{\partial\psi}{\partial t}=\left(  -\frac{\hbar^{2}}{2M}\nabla
^{2}+V_{\bot}(\mathbf{r}_{\bot})+V_{z}(z)+gN\left\vert \psi\right\vert
^{2}\right)  \psi, \label{II-0a}%
\end{equation}
which has proven to be very successful in reproducing experimental results for
the zero-temperature BEC (for instance, for multiple vortices, as shown in
detail in Refs. \cite{Vort,Huht}). In this equation, $M$ is the atomic mass,
the norm of the wave function $\psi$ is unity, $N$ is the number of atoms, and
$g=4\pi\hbar^{2}a/M$ is the interaction strength with $a$ being the
\textsl{s}-wave scattering length. In this work, we consider only repulsive
interactions (i.e., $a>0$). Furthermore, $V_{\bot}(\mathbf{r}_{\bot
})=(1/2)M\omega_{\bot}^{2}r_{\bot}^{2}$ is the radial-confinement potential
and $V_{z}(z)$ is the axial potential, which may include the axial harmonic
trap and the 1D OL, with depth $V_{0}$ and period $d$:
\begin{equation}
V_{z}(z)=(1/2)M\omega_{z}^{2}z^{2}+V_{0}\sin^{2}\left(  \pi z/d\right)  .
\label{II-0b}%
\end{equation}
The energy scale in the underlying (no-interaction) linear problem is set by
the OL's recoil energy, $E_{R}=\hbar^{2}\pi^{2}/(2Md^{2})$.

Most commonly, theoretical studies of GSs in elongated geometries are carried
out in terms of the 1D GPE \cite{Kon1,Mor1,Kiv1,Abd1,Mal1,Wu1,Wu2}:%
\begin{equation}
i\hbar\frac{\partial\phi}{\partial t}=-\frac{\hbar^{2}}{2M}\frac{\partial
^{2}\phi}{\partial z^{2}}+V_{z}(z)\phi+g_{\mathrm{1D}}N\left\vert
\phi\right\vert ^{2}\phi, \label{II-0c}%
\end{equation}
where $g_{\mathrm{1D}}=2a\hbar\omega_{\bot}$. This is an effective evolution
equation for the axial dynamics ---described by the 1D wave function
$\phi(z,t)$--- which can be derived from the full three-dimensional (3D) GPE
after averaging out the radial degrees of freedom under the assumption that
the radial confinement is so tight that the transverse dynamics are frozen to
zero-point oscillations. These conditions, however, are not easy to realize
using typical experimental parameters \cite{Mor1}, and when such conditions
are not met the transverse excitations can no longer be neglected, making an
essentially 3D analysis necessary. In this work, we aim to investigate
different physically relevant regimes and capture 3D effects in the generation
and stability of matter-wave GSs in elongated BECs. This analysis should make
it possible to determine the range of validity of the 1D GPE for the
description of the GSs under typical experimental conditions, as well as
characteristic features of the fundamental and higher-order GSs in realistic situations.

In principle, when the transverse excitations are relevant, Eq. (\ref{II-0c})
fails and one has to resort to the full 3D GPE (\ref{II-0a}), as recently done
in Ref. \cite{PRA82}, or, alternatively, use extended 1D models that, within
the framework of certain assumptions, take into account effects of
higher-order radial modes on the axial dynamics of the condensate, leading to
effective 1D equations with the cubic-quintic \cite{CQ} or nonpolynomial
\cite{Reatto1,PRA77} nonlinearities. In this work, we will consider both the
full 3D GPE and the effective 1D model with a nonpolynomial nonlinearity,
which was derived, for the case of the self-repulsive nonlinearity, in Ref.
\cite{PRA77}. The latter one, which represents a simple generalization of the
usual 1D GPE, that reduces to Eq. (\ref{II-0c}) in the appropriate limit, has
demonstrated an excellent quantitative agreement with experimental
observations \cite{Kev1,Wel1,Kev2,Carre3} (chiefly, for delocalized dark
solitons, which are natural patterns in the case of the self-repulsion;
however, the comparison was not reported before for localized GS modes). We
will demonstrate that, while the range of applicability of the 1D GPE
(\ref{II-0c}) is severely limited in realistic situations, the above-mentioned
generalization gives a good description of stationary matter-wave GSs in
virtually all cases of practical interest.

The paper is organized as follows. The model is formulated in Section II,
where we also recapitulate the derivation of the effective nonpolynomial 1D
equation following the lines of Ref. \cite{PRA77}, as this derivation is
essential for the presentation of the results for the GSs. The main findings
are collected in Sections III, where families of GS solutions are reported in
several physically relevant regimes for strong, intermediate, and weak
transverse confinement and shallow or deep OL potential. Both fundamental GSs
and their multipeak bound states are considered. Only in the case of tight
confinement combined with a shallow OL does the ordinary 1D GPE provide for a
sufficiently accurate description of the GSs in comparison with results of
full 3D computations. On the other hand, the effective nonpolynomial 1D
equation provides for good accuracy in all cases; for the fundamental solitons
and their bound states alike. The stability of the GSs is studied in Section
IV by means of systematic simulations of the evolution of perturbed solitons.
The conclusion is that the fundamental GSs are stable (except in a narrow
region close to the upper edge of the band gaps), even in the case of strong
deviation from the usual 1D description. Multisoliton bound states are stable
if the OL potential is deep enough. Conclusions following from results of this
work, including applicability limits for the mean-field approximation and 1D
approximations, are formulated in Section V.

\section{THE MODEL}

The model that was developed in Ref. \cite{PRA77} for a BEC in the absence of
OL potentials resorted to the adiabatic approximation, to neglect correlations
between the transverse and axial motions. This approximation assumes that the
axial density varies slowly enough to allow the transverse wave function to
follow these slow variations. Because the OL imposes a new spatial scale,
which may be more restrictive, it is necessary to find out if the adiabatic
approximation remains valid in the presence of the OL. To this end, we will
now briefly recapitulate the derivation of the effective 1D equation based on
this approximation.

The starting point is the ansatz based on the factorized 3D wave function%
\begin{equation}
\psi(\mathbf{r},t)=\varphi(\mathbf{r}_{\bot};n_{1}(z,t))\phi(z,t),
\label{II-1}%
\end{equation}
with $n_{1}(z,t)$ being the local condensate density per unit length
characterizing the axial configuration:%
\begin{equation}
n_{1}(z,t)\equiv N\int d^{2}\mathbf{r}_{\bot}|\psi(\mathbf{r}_{\bot}%
,z,t)|^{2}=N|\phi(z,t)|^{2}. \label{II-2}%
\end{equation}
To derive Eq. (\ref{II-2}), we have assumed that the transverse wave function
$\varphi$ is normalized to unity. Next, the substitution of Eq. (\ref{II-1})
into the 3D equation (\ref{II-0a}) leads to%
\begin{gather}
\left(  i\hbar\frac{\partial\phi}{\partial t}+\frac{\hbar^{2}}{2M}%
\frac{\partial^{2}\phi}{\partial z^{2}}-V_{z}\phi\right)  \varphi=\nonumber\\
\left(  \!\!-\frac{\hbar^{2}}{2M}\!\nabla_{\!\bot}^{2}\varphi-\!\frac
{\hbar^{2}}{2M}\frac{\partial^{2}\varphi}{\partial z^{2}}\!+\!V_{\!\bot
}\varphi\!+\!gn_{1}\!\left\vert \varphi\right\vert ^{2}\!\!\varphi\!\!\right)
\!\phi-\frac{\hbar^{2}}{M}\frac{\partial\varphi}{\partial z}\frac{\partial
\phi}{\partial z}. \label{II-5}%
\end{gather}
Because of the very different time scales of the axial and radial motions, it
is natural to assume that the slow axial dynamics may be accurately described
by averaging Eq. (\ref{II-5}) over the fast (radial) degrees of freedom. Doing
this, one obtains%
\begin{equation}
i\hbar\frac{\partial\phi}{\partial t}\!=\!-\frac{\hbar^{2}}{2M}\frac
{\partial^{2}\phi}{\partial z^{2}}+V_{z}\phi+\mu_{\bot}\phi-\frac{\hbar^{2}%
}{M}\!\left(  \!\int\!\!d^{2}\mathbf{r}_{\bot}\varphi^{\ast}\frac
{\partial\varphi}{\partial z}\!\right)  \!\frac{\partial\phi}{\partial z},
\label{II-6}%
\end{equation}%
\begin{equation}
\mu_{\bot}\!(n_{1}\!)\!\equiv\!\!\!\int\!\!d^{2}\mathbf{r}_{\bot}\varphi
^{\ast}\!\!\left(  \!\!-\frac{\hbar^{2}}{2M}\nabla_{\!\bot}^{2}\!-\!\frac
{\hbar^{2}}{2M}\frac{\partial^{2}}{\partial z^{2}}\!+\!V_{\!\bot}%
\!+\!gn_{1}\!\left\vert \varphi\right\vert ^{2}\!\!\right)  \!\varphi.
\label{II-7}%
\end{equation}
Since $n_{1}(z,t)$ enters the last term of Eq. (\ref{II-7}) merely as an
external parameter, it is clear that, whenever the axial kinetic energy
associated with the transverse wave function may be neglected; namely,%
\begin{align}
K_{z}[\varphi]  &  \equiv\int\!d^{2}\mathbf{r}_{\bot}\varphi^{\ast}\!\left(
\!-\frac{\hbar^{2}}{2M}\frac{\partial^{2}}{\partial z^{2}}\right)
\varphi\nonumber\\
&  \ll\int\!d^{2}\mathbf{r}_{\bot}\varphi^{\ast}\!\left(  \!-\frac{\hbar^{2}%
}{2M}\nabla_{\bot}^{2}+V_{\bot}(\mathbf{r}_{\bot})+gn_{1}\left\vert
\varphi\right\vert ^{2}\!\right)  \!\varphi, \label{II-9}%
\end{align}
the radial dynamics decouple and $\mu_{\bot}(n_{1})\!$ can be determined
without the knowledge of the axial wave function. Actually, the right-hand
side of Eq. (\ref{II-9}) is the chemical potential of an axially homogeneous
condensate characterized by the density per unit length $n_{1}$ and wave
function $\varphi$. For a sufficiently small value of the linear density
($an_{1}\ll1$), the chemical potential of the lowest-energy state of this
homogeneous condensate can be readily obtained perturbatively. In this case,
to the lowest order, $\varphi(\mathbf{r}_{\bot})$ is given by the Gaussian
wave function of the ground state of the radial harmonic oscillator, with the
corresponding chemical potential being $\hbar\omega_{\bot}(1+2an_{1})$. As the
linear density increases, more radial modes become excited and, in general,
the ground state of the corresponding homogeneous condensate involves many
harmonic-oscillator modes.

In previous works, it was shown using different approaches that, also in this
case, an analytical solution can be constructed \cite{PRA77}. In particular,
by using a variational approach based on the direct minimization of the
chemical-potential functional, it was shown that, for any (dimensionless)
linear density $an_{1}$, a very accurate estimate for the chemical potential
of the ground state is given by $\hbar\omega_{\bot}\sqrt{1+4an_{1}}$
\cite{Ger1}. This expression can be easily extended to incorporate the case in
which the condensate contains an axisymmetric vortex \cite{PRA77} (see also
Ref. \cite{Luca}, where the intrinsic vorticity was included into the
derivation of the 1D nonpolynomial nonlinear Schr\"{o}dinger equation in the
case of self-attraction). However, in this work we restrict the consideration
to zero vorticity. Using Eq. (\ref{II-9}), we see that a sufficient condition
for the second derivative in $z$ appearing in Eq. (\ref{II-7}) to be
negligible is%
\begin{equation}
K_{z}[\varphi]\ll\hbar\omega_{\bot}\sqrt{1+4an_{1}}. \label{II-10}%
\end{equation}
Taking into account an estimate, $K_{z}[\varphi]\sim\hbar^{2}/(2M\Delta
_{z}^{2})$, where $\Delta_{z}$ is the characteristic length scale in the axial
direction, Eq. (\ref{II-10}) can be rewritten as%
\begin{equation}
\frac{\hbar^{2}}{2M\Delta_{z}^{2}}\ll\hbar\omega_{\bot}\sqrt{1+4an_{1}}.
\label{II-10b}%
\end{equation}
When this condition holds, $\mu_{\bot}(n_{1})$ coincides, to a good
approximation, with the transverse local chemical potential of the stationary
radial GPE:%
\begin{equation}
\left(  \!-\frac{\hbar^{2}}{2M}\nabla_{\bot}^{2}+V_{\bot}(\mathbf{r}_{\bot
})+gn_{1}\left\vert \varphi\right\vert ^{2}\!\right)  \varphi=\mu_{\bot}%
(n_{1})\varphi, \label{II-11}%
\end{equation}
and is given by%
\begin{equation}
\mu_{\bot}(n_{1})=\hbar\omega_{\bot}\sqrt{1+4an_{1}}. \label{II-12}%
\end{equation}

Finally, substituting Eqs. (\ref{II-12}) and (\ref{II-2}) into Eq.
(\ref{II-6})\ and taking into account that $\varphi$ is a real normalized wave
function (which implies the vanishing of the integral in the last term), we
arrive at the following effective 1D equation to govern the slow axial
dynamics of the condensate:%
\begin{equation}
i\hbar\frac{\partial\phi}{\partial t}=-\frac{\hbar^{2}}{2M}\frac{\partial
^{2}\phi}{\partial z^{2}}+V_{z}(z)\phi+\hbar\omega_{\bot}\sqrt{1+4aN|\phi
|^{2}}\phi.\label{II-13}%
\end{equation}
We note that the contribution from interatomic interactions enters the above
equation through term $\hbar\omega_{\bot}(\sqrt{1+4aN|\phi|^{2}}-1)\phi$,
which vanishes for $N\rightarrow0$. Thus, Eq. (\ref{II-13}) incorporates an
energy shift $\hbar\omega_{\bot}$, which is irrelevant for the dynamics but
simplifies the form of the equation and makes the global chemical potential
that follows from this effective 1D equation exactly coinciding with the
corresponding 3D result.

The above derivation demonstrates that the validity of Eq. (\ref{II-13})
relies on two conditions:

(i) The typical time scale $\Delta_{t}$ of the axial motion must be much
larger than the typical time scale of the radial motion ($\sim\omega_{\bot
}^{-1}$).

(ii) The axial kinetic energy $K_{z}[\varphi]$ associated with the transverse
wave function must be negligible.

The former requirement is necessary to allow the radial wave function to
adiabatically follow the axial dynamics. While the specific temporal scale
$\Delta_{t}$ depends on the particular initial conditions, typically, in the
presence of an OL, the fulfillment of this condition gets more difficult as
the lattice period $d$ decreases, or the linear density $an_{1}$ increases. In
particular, a sufficient condition for the validity of the adiabatic
approximation is: $d\gg a_{\bot}$, $an_{1}\ll1$, with $a_{\bot}\equiv
\sqrt{\hbar/(M\omega_{\bot})}$ being the radial harmonic-oscillator length.
Nevertheless, such a constraint, which is hard to satisfy in realistic
situations, is not a necessary condition. Actually, for stationary states
$\Delta_{t}\rightarrow\infty$ and condition (i) always holds. In the present
work we are interested in this case, which is the most relevant one to seek
for matter-wave GSs.

A sufficient condition for the fulfillment of condition (ii) is given by the
inequality (\ref{II-10b}). In the presence of an OL, the characteristic length
scale $\Delta_{z}$ is typically on the order of the lattice period $d $, hence
Eq. (\ref{II-10b}) becomes%
\begin{equation}
E_{R}\ll\hbar\omega_{\bot}\sqrt{1+4an_{1}}, \label{II-14}%
\end{equation}
where $E_{R}$ is the corresponding recoil energy. We will demonstrate that
this condition is well satisfied for stationary GSs in most cases of interest.

It is clear that, when the linear density is small enough ($an_{1}\ll1$), Eq.
(\ref{II-13}), which exhibits the nonpolynomial nonlinearity, reduces to the
usual 1D GPE (\ref{II-0c}) with the cubic nonlinearity. This is the quasi-1D
mean-field regime, which corresponds to condensates whose radial state, as
given by a solution to Eq. (\ref{II-11}), may be well approximated by the
Gaussian ground state of the corresponding harmonic oscillator
\cite{PRA77,PRA74}. Thus, Eq. (\ref{II-13}) represents an extension of the 1D
GPE for condensates with larger linear densities or, equivalently, larger
mean-field interaction energies. In such condensates, the radial ground state
satisfying Eq. (\ref{II-11}) is, in general, a linear combination of many
harmonic-oscillator modes (a similar situation takes place in the derivation
of the effective 1D equation for fermionic gases by means of the
density-functional approach \cite{SKA}).

Inspection of Eqs. (\ref{II-0a}) or (\ref{II-13}) demonstrates that the
dynamical problem is fully controlled by four dimensionless parameters, which
may be defined as%
\begin{equation}
\frac{E_{R}}{\hbar\omega_{\bot}},\;\frac{V_{0}}{E_{R}},\;\frac{\omega_{z}%
}{\omega_{\bot}},\;\frac{Na}{a_{\bot}}.
\end{equation}
As said above, the recoil energy $E_{R}$ sets the energy scale of the
underlying linear problem, while the nonlinear coupling constant $Na/a_{\bot}$
determines the order of magnitude of the mean-field interaction energy in
units of the radial quantum $\hbar\omega_{\bot}$. Note that the dimensionless
linear density $an_{1}$ is proportional to $Na/a_{\bot}$.

In this work, we assume the axial confinement to be so weak that the
corresponding harmonic-oscillator length, $a_{z}\equiv\sqrt{\hbar/(M\omega
_{z})}$, is much larger than the lattice period $d$. Under these conditions,
it is safe to neglect the modulation induced in the condensate density by the
axial harmonic trap and set $\omega_{z}=0$, so that we are left with a uniform
1D lattice potential acting along the axial direction.

\section{MATTER-WAVE GAP SOLITONS IN DIFFERENT PHYSICAL REGIMES}

The stationary solutions of the 3D GPE are wave functions of the form
$\psi(\mathbf{r},t)=\psi_{0}(\mathbf{r})\exp(-i\mu t)$, with $\psi_{0}$
obeying the time-independent GPE:%
\begin{equation}
\mu\psi_{0}=\left(  -\frac{\hbar^{2}}{2M}\nabla^{2}+V_{\bot}(\mathbf{r}_{\bot
})+V_{z}(z)+gN\left\vert \psi_{0}\right\vert ^{2}\right)  \psi_{0}%
,\label{II-17}%
\end{equation}
where $\mu$ is the chemical potential. When Eq. (\ref{II-13}) is applicable,
one can instead generate the stationary axial wave function $\phi_{0}(z)$ by
solving the effective 1D equation%
\begin{equation}
\mu\phi_{0}=\left(  -\frac{\hbar^{2}}{2M}\frac{\partial^{2}}{\partial z^{2}%
}+V_{z}(z)+\hbar\omega_{\bot}\sqrt{1+4aN|\phi_{0}|^{2}}\right)  \phi
_{0},\label{II-18}%
\end{equation}
which, in the limit of $an_{1}\ll1$, reduces to the time-independent version
of the usual 1D GPE (\ref{II-0c}):%
\begin{equation}
\mu\phi_{0}=\left(  -\frac{\hbar^{2}}{2M}\frac{\partial^{2}}{\partial z^{2}%
}+V_{z}(z)+g_{\mathrm{1D}}N\left\vert \phi_{0}\right\vert ^{2}+\hbar
\omega_{\bot}\right)  \phi_{0}.\label{II-18b}%
\end{equation}

Matter-wave GSs are characterized by a chemical potential lying in a band gap
of the energy spectrum of the underlying linear system. To construct solutions
for realistic 3D matter-wave GSs in the effectively 1D geometry, we looked for
numerical solutions to Eqs. (\ref{II-17}) and (\ref{II-18}) in different
physically relevant regimes, and compared the obtained results with those
produced by the usual 1D GPE (\ref{II-18b}) to determine to what extent 3D
contributions are relevant. In particular, the comparison allows us to
estimate the accuracy and range of applicability of the effective 1D equations
(\ref{II-18}) and (\ref{II-18b}).

Experimentally relevant values for the OL period $d$ range from $0.4$ to $1.6
$ $\operatorname{\mu m}$ \cite{Mor1}, which implies that, for the condensate
of $^{87}$Rb, $E_{R}/(2\pi\hbar)$ ranges from $3.6$ $\operatorname{kHz}$ to
$220$ $\operatorname{Hz}$. Taking into account that, typically, $\omega_{\bot
}/2\pi\lesssim1$ $\operatorname{kHz}$, we conclude that $E_{R}/\hbar
\omega_{\bot}\gtrsim1/4$. On the other hand, for the condensate of $^{23}$Na
one has $E_{R}/\hbar\omega_{\bot}\gtrsim1$ \cite{Na}; therefore, in this case
the GS energy is sufficiently large to excite higher modes of the radial
confinement, which makes 3D contributions always relevant. Since the $^{87}$Rb
condensate is most relevant for the experimental realization of GSs in the
quasi-1D setting \cite{Ober1,Isa1}, in what follows below we use particular
parameters of this condensate to illustrate our results. Nevertheless, the
results of the present work, which are always expressed in terms of relevant
dimensionless parameters, are valid for any BEC. Actually, this is a direct
consequence of the scaling properties of Eqs. (\ref{II-0a}), (\ref{II-0c}),
and (\ref{II-13}).

\subsection{Tight radial confinement: $E_{R}/\hbar\omega_{\bot}\ll1$}

In this case, the quantum of radial excitations is much greater than the
typical energy scale in the linear problem. Note that, to realize this regime
in a $^{87}$Rb condensate, even in an OL of period $d\simeq1.6$
$\operatorname{\mu m}$, one needs to use a harmonic trap with radial
frequency $\omega_{\bot}/2\pi\gtrsim2400$ $\operatorname{Hz}$.


\begin{figure}[ptb]
\begin{center}
\includegraphics[
width=8.2cm
]{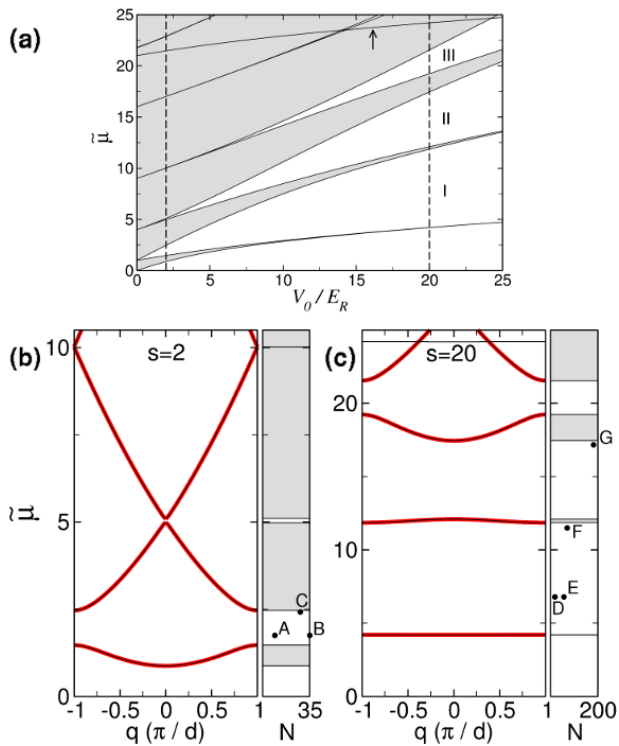}
\end{center}
\caption{(Color online) (a) Band-gap structure of a noninteracting 3D BEC
with $E_{R}/\hbar\omega_{\bot}=1/10$, as a function of the dimensionless
lattice depth $s\equiv V_{0}/E_{R}$. The representative cases $s=2$ and $20$,
indicated by vertical dashed lines, are considered in more detail in (b) and
(c), respectively, which show the dimensionless chemical potential
$\widetilde{\mu}$ as a function of the quasimomentum $q$ in the first
Brillouin zone (in units of $\pi/d$). The right panels display the location of
the $^{87}$Rb gap solitons considered in this work (points A--G).}%
\label{Fig1}%
\end{figure}


Figure \ref{Fig1}(a) shows the dimensionless chemical potential of the
noninteracting 3D condensate with $E_{R}/\hbar\omega_{\bot}=1/10$,%
\begin{equation}
\widetilde{\mu}\equiv(\mu-\hbar\omega_{\bot})/E_{R}, \label{II-18c}%
\end{equation}
as a function of the dimensionless lattice depth, $s\equiv V_{0}/E_{R}$. Since
in this work we are interested in GSs with zero vorticity, this diagram, which
represents the band-gap structure of the underlying linear problem, has been
obtained from the zero-vorticity solutions of the linear version of Eq.
(\ref{II-17}). Regions I, II, and III correspond to the lowest finite
band gaps, which separate shaded bands, where linear solutions exist. An
important point is that, within the region of interest in the parameter space
(for $V_{0}/E_{R}\lesssim25$ and up to the third band gap), this 3D diagram is
indistinguishable from that obtained using the linear version of the effective
1D equation (\ref{II-18}) [which, obviously, coincides with the linear version
of the stationary 1D GPE (\ref{II-18b})]. In fact, Eq. (\ref{II-18}) leads to
a band-gap diagram that differs from Fig. \ref{Fig1}(a) solely in the band
marked by the arrow, which does not appear in the 1D case. This extra band is,
essentially, a replica of the lowest one, shifted up in energy by
$2\hbar\omega_{\bot}/E_{R}$. Taking into account that the energy spectrum of
the radial harmonic oscillator is given by%
\begin{equation}
E=(2n_{r}+|m|+1)\hbar\omega_{\bot}, \label{II-19}%
\end{equation}
where $n_{r}=0,1,2,\ldots$ is the radial quantum number and $m=0,\pm
1,\pm2,\ldots$ is the axial angular-momentum quantum number, it is evident
that the additional up-shifted band corresponds to the first-excited radial
mode with $m=0$. In the notation of Ref. \cite{PRA82}, it corresponds to
quantum numbers $(n=1,m=0,n_{r}=1)$, with $n$ being the band index of the 1D
axial problem. Thus, the appearance of this band is a purely 3D effect that
cannot be accounted for by the above 1D models. Since the energy shift
increases as $E_{R}/\hbar\omega_{\bot}$ decreases, it is clear that, within
the parameter region of interest, Fig. \ref{Fig1}(a) is universal in the sense
that it is valid (for both 1D and 3D systems) for any 
$E_{R}/\hbar\omega_{\bot}\ll1$.

In this work, we restrict ourselves to the case of $(1,0,0)$ GSs; that is, the
solitons bifurcating from the lowest-energy band. Three-dimensional
matter-wave GSs with a nontrivial radial structure have been studied in Ref.
\cite{PRA82}. In this subsection, we consider OLs with $E_{R}/\hbar
\omega_{\bot}=1/10$ and depth $V_{0}/E_{R}=2$ or $20$. These two
representative cases correspond to the dashed vertical lines in Fig.
\ref{Fig1}(a) and are shown in more detail in Figs. \ref{Fig1}(b) and 
\ref{Fig1}(c), which display the corresponding band-gap structure as a function 
of the quasimomentum in the first Brillouin zone (in units of $\pi/d$). Bold 
red lines in Figs. \ref{Fig1}(b) and \ref{Fig1}(c) have been obtained from the 
linear version of the effective 1D equation (\ref{II-18}) while thin black lines
correspond to results obtained by means of the linear version of the 3D
equation (\ref{II-17}) that cannot be reproduced with the 1D model. As mentioned
above, the only appreciable difference between the 1D and 3D results comes
from the contribution of the first-excited radial mode [see the top part of
Fig. \ref{Fig1}(c)]. The case of $V_{0}/E_{R}=2$, displayed in Fig.
\ref{Fig1}(b), corresponds to the condensate in relatively shallow OL
potentials. Under these circumstances, the linear energy spectrum does not
differ too much from that corresponding to the translationally invariant case
($V_{z}=0$), and the band-gap structure exhibits wide energy bands separated by
relatively small gaps, which are tangible only in the lowest part of the
spectrum. On the contrary, for $V_{0}/E_{R}=20$ [the tight-binding regime, 
see Fig. \ref{Fig1}(c)] the condensate is trapped in the deep periodic
potential. In this case, the lowest part of the linear spectrum is dominated
by large gaps separating relatively narrow energy bands. The panels on the
right side of Figs. \ref{Fig1}(b) and \ref{Fig1}(c) show the location, with 
respect to the corresponding linear band-gap diagram, of the GSs that will be 
considered below (points A--G). The horizontal axes in these panels indicate the 
number of atoms in each GS for the $^{87}$Rb condensates (with the \textsl{s}-wave
scattering length $a=5.29\operatorname{nm}$) in the trap with radial frequency
$\omega_{\bot}/2\pi=2400$ $\operatorname{Hz}$. For other parameter values,
the GS family is described by the same plots, if considered in terms of the
above-mentioned nonlinear coupling constant, $Na/a_{\perp}$
[$a=5.29\operatorname{nm}$ and $\omega_{\bot}/2\pi=2400$ $\operatorname{Hz}$
correspond to $a/a_{\bot}=0.024$]. In other words, the number of atoms in a GS
created in the condensate with scattering length $a^{\prime}$ and transverse
confinement radius $a_{\perp}^{\prime}$, which is tantamount to the GS with
particular values of $a$, $a_{\perp}$ and $N$, is given by%
\begin{equation}
N^{\prime}=\frac{aa_{\bot}^{\prime}}{a^{\prime}a_{\bot}}N. \label{II-19c}%
\end{equation}

GS solutions have been obtained by numerically solving the full 3D equation
(\ref{II-17}), as well as the effective 1D equations (\ref{II-18}) and
(\ref{II-18b}). To this end, we have implemented a Newton continuation
method, based on a Laguerre--Fourier spectral basis, that uses the chemical
potential $\mu$ as a continuation parameter. To ensure the convergence,
methods of this kind require initiation of the iterative procedure with a
sufficiently good initial guess. This means that one needs a good estimate for
both the axial and the radial parts of the wave function. Our computations
used the following initial ansatz for the wave function:%
\begin{equation}
\psi_{0}(r_{\bot},z)=\frac{1}{\Gamma a_{\bot}\sqrt{\pi}}\exp\left(
-\frac{r_{\bot}^{2}}{2\Gamma^{2}a_{\bot}^{2}}\right)  \phi_{0}(z),
\label{II-20}%
\end{equation}
where the $z$-dependent condensate's width, expressed in units of $a_{\bot}$,
is%
\begin{equation}
\Gamma=\left(  1+2aN|\phi_{0}(z)|^{2}\right)  ^{1/4}, \label{II-21}%
\end{equation}
and $\phi_{0}(z)=\sqrt{k_{0}/2}\;\mathrm{sech}(k_{0}z)$ is the axial wave
function. The number of particles, $N$, and $k_{0}$ are free adjustable
parameters. Note that the radial part of the wave function (\ref{II-20}) is
the same as the variational solution of Eq. (\ref{II-11}), which was found in
Ref. \cite{PRA77}.


\begin{figure}[ptb]
\begin{center}
\includegraphics[
width=8.2cm
]{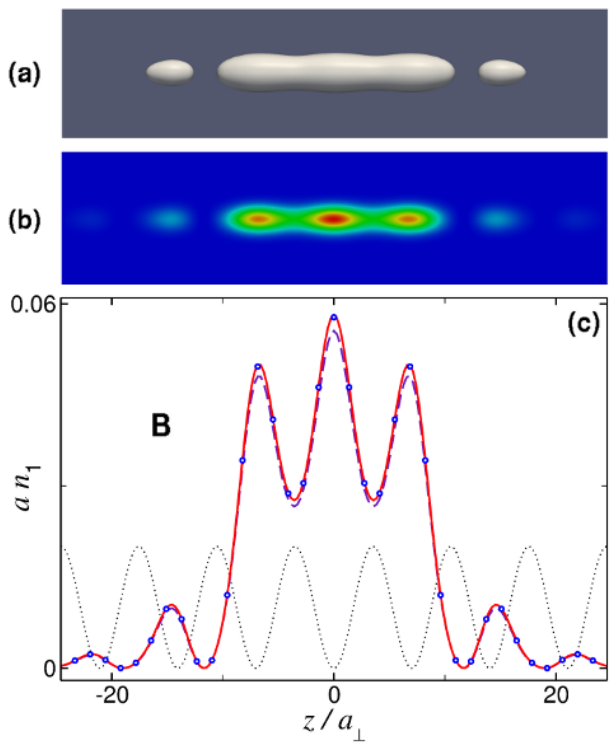}
\end{center}
\caption{(Color online) Atom density of the gap soliton corresponding to
point B in Fig. \ref{Fig1}(b), displayed as (a) an isosurface taken at $5\%$
of the maximum density and (b) as a color map along a cutting plane
containing the $z$ axis. (c) Dimensionless axial density $an_{1}$ obtained
from the 3D wave function, as prescribed by Eq. (\ref{II-2}) (open circles)
along with the corresponding predictions from the nonpolynomial 1D equation
(\ref{II-18}) (solid red line) and the ordinary 1D GPE (\ref{II-18b}) (dashed
blue line).}%
\label{Fig2}%
\end{figure}


\subsubsection{Shallow optical lattice: $V_{0}/E_{R}=2$}

Point A in Fig. \ref{Fig1}(b) corresponds to a fundamental GS located near the
bottom edge of the first band gap. It looks qualitatively similar to that shown
below in Fig. \ref{Fig3}, but with the peak linear density $an_{1}%
(0)\simeq0.04$. In this case, the effective 1D equations (\ref{II-18}) and
(\ref{II-18b}) yield results which are indistinguishable, on the scale of the
figures, from those obtained from the full 3D equation (\ref{II-17}). This is
not surprising because, for these parameters, condition (\ref{II-14}) is certainly
satisfied, and inequality $an_{1}\ll1$, which guarantees the validity of the
1D GPE (\ref{II-18b}), also holds. The (dimensionless) chemical potential of
this GS is $\widetilde{\mu}=1.75$, which implies $\mu=1.175\,\hbar\omega
_{\bot}\simeq\hbar\omega_{\bot}$, hence the radial wave function of the
condensate should not differ too much from the ground state of the
corresponding harmonic oscillator. These fundamental GSs, however, can only
accommodate $9$ particles (for the $^{87}$Rb condensate), which is too small
to use the mean-field approximation. Yet these solutions play an
important role as building blocks of higher-order GSs. Point B in Fig.
\ref{Fig1}(b) corresponds to one of these higher-order solitons, which is
built as a symmetric linear combination of three fundamental GSs, see Fig.
\ref{Fig2}. Its chemical potential and number of particles are $\widetilde
{\mu}=1.75$ and $N=35$. Note that, for the relatively shallow OL ($V_{0}%
/E_{R}=2$), interference effects arising from the overlap between fundamental
GSs sitting in adjacent lattice cells play an important role. For this reason,
the contrast between the three central peaks and minima separating them is
rather low in Fig. \ref{Fig2}, and the total number of particles in the
compound soliton does not coincide with the sum of the numbers of its
fundamental constituents. Extending this procedure, one can readily build a
sequence of compound GSs with an increasing number of peaks.


\begin{figure}[ptb]
\begin{center}
\includegraphics[
width=8.2cm
]{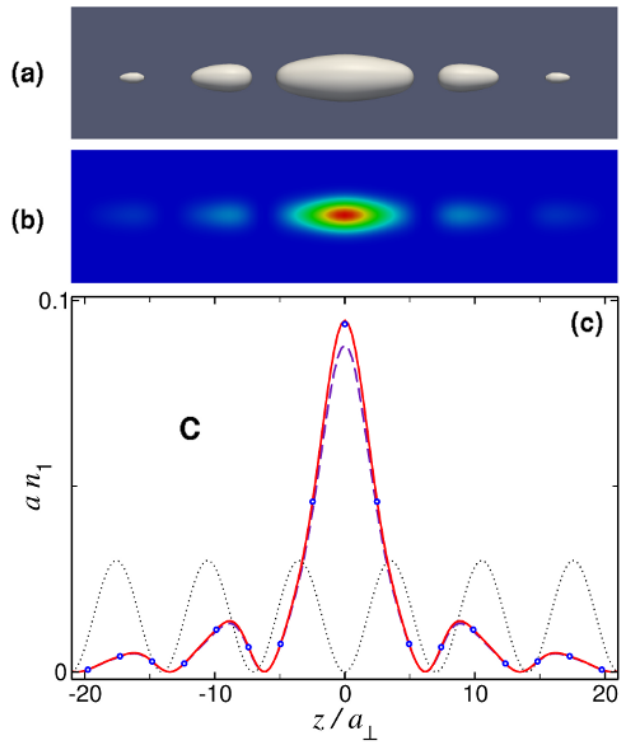}
\end{center}
\caption{(Color online) Same as Fig. \ref{Fig2} but for the gap soliton
corresponding to point C in Fig. \ref{Fig1}(b).}%
\label{Fig3}%
\end{figure}


Panels (a) and (b) in Fig. \ref{Fig2} display the 3D density of the
three-peak GSs as, respectively, an isosurface (taken at $5\%$ of the
maximum density) and a color map in the cross-section plane drawn through the
$z$ axis. From these results, which have been obtained by solving the full 3D
equation (\ref{II-17}), we have also derived the axial linear density
$n_{1}(z)$, defined as per Eq. (\ref{II-2}). This density is shown in Fig.
\ref{Fig2}(c) by open circles, along with the corresponding results obtained
from the effective 1D equations (\ref{II-18}) and (\ref{II-18b}). The OL
potential is also displayed by the dotted line (in arbitrary units). It is
seen that the effective 1D equation (\ref{II-18}) (solid red lines) reproduces
the 3D results very accurately. The 1D GPE (\ref{II-18b}) (dashed blue lines)
gives a slight discrepancy ($\simeq3\%$) against the 3D results. This
discrepancy, which would be invisible in the isolated fundamental GSs that
build the compound, may also be explained by effects of the interference
between the constituents.

Point C in Fig. \ref{Fig1}(b) indicates the location of a fundamental gap
soliton in the $\left(  N,\widetilde{\mu}\right)  $ plane, which sits near the
top edge of the first band gap. It contains $28$ particles and has chemical
potential $\widetilde{\mu}=2.42$, which corresponds to $\mu=1.242\,\hbar
\omega_{\bot}$. This quantity is again very similar to $\hbar\omega_{\bot}$,
so that one expects the 1D GPE to be a good approximation in this case. Figure
\ref{Fig3} shows the 3D density and the axial linear density $an_{1}(z)$ of
this GS. In Fig. \ref{Fig3}(c) one can see that the 1D GPE (\ref{II-18b}) (the
dashed blue line) reproduces the 3D results obtained from the full GPE
(\ref{II-17}) (shown by open circles) within a $5\%$ deviation. The effective
1D equation (\ref{II-18}) (whose results are represented by the solid red
line) is again more accurate and reproduces the 3D results with an error
$<1\%$. As before, this fundamental GS can be used to build multipeak compounds.

As one moves upward in the bandgap, the 3D effects become stronger, which is a
consequence of the fact that the corresponding number of particles and, thus,
the nonlinear interaction energy increase. The above results, however,
indicate that, for the tight radial confinement ($E_{R}/\hbar\omega_{\bot}%
\ll1$) and shallow OL ($V_{0}/E_{R}\leq2$), the specific 3D effects may 
eventually be neglected. In fact, under these conditions the maximum number of
particles that can be accommodated in a fundamental GS in the first band gap is
so small that inequality $an_{1}\ll1$ always holds. This means that both the
linear and the nonlinear energies remain much smaller than the quantum of the
radial excitation ($E_{R},\,an_{1}\hbar\omega_{\bot}\ll\hbar\omega_{\bot}$),
hence no higher transverse modes are significantly excited. Therefore, the
situation considered in this subsection may be categorized as the quasi-1D
mean-field regime, in which the 1D GPE (\ref{II-18b}) accurately describes the
matter-wave GSs, provided that condition $N\gg1$ holds (otherwise, the
mean-field approximation will be invalid).

\subsubsection{Deep optical lattice: $V_{0}/E_{R}=20$}

\begin{figure}[ptb]
\begin{center}
\includegraphics[
width=8.2cm
]{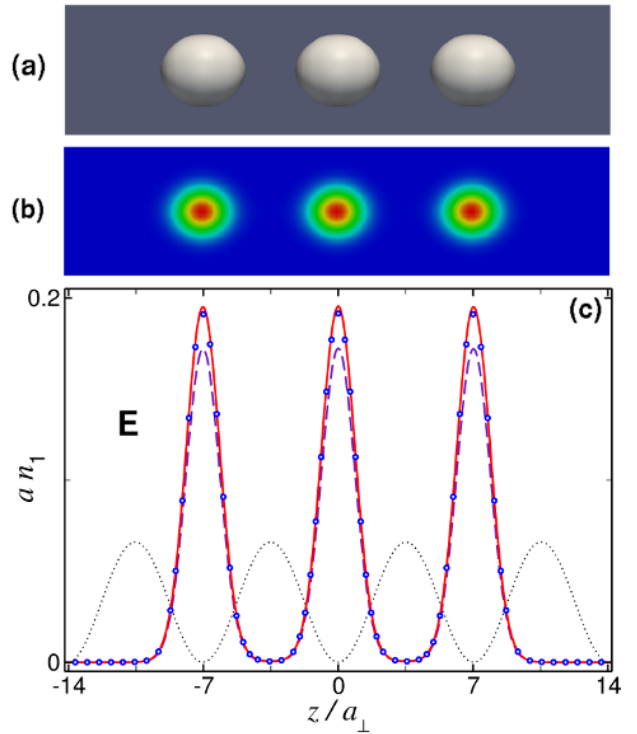}
\end{center}
\caption{(Color online) Same as Fig. \ref{Fig2} but for the gap soliton
corresponding to point E in Fig. \ref{Fig1}(c).}%
\label{Fig4}%
\end{figure}

In terms of the underlying OL, this is the case of the tight-binding regime,
$V_{0}\gg E_{R}$. The nonlinear tight-binding regime is realized when the
potential depth $V_{0}$ is much larger than both the linear and nonlinear
(mean-field) energies (i.e.,$\ V_{0}\gg E_{R},\,an_{1}\hbar\omega_{\bot}$).
Under these circumstances, the condensate density is highly localized near
potential minima, making the overlap between densities associated with
different wells negligible. Point D in the first band gap of Fig. \ref{Fig1}(c)
corresponds to a fundamental GS with $\widetilde{\mu}=6.79$ and $N=18$. Since
its peak axial density is $an_{1}(0)\simeq0.2$, and $V_{0}/E_{R}=20$ implies
$V_{0}=2\,\hbar\omega_{\bot}$, this fundamental GS belongs to the nonlinear
tight-binding regime. Figure \ref{Fig4} shows a compound GS\ built of three
such fundamental solitons. It corresponds to point E in Fig. \ref{Fig1}(c).
Its chemical potential, $\widetilde{\mu}=6.79$, is the same as that of the
corresponding fundamental GS, and its number of particles, $N=55$, now
coincides with the sum of the number in its constituents (note that we always
approximate $N$ to the nearest integer). As can be seen from Fig. \ref{Fig4},
this compound soliton exhibits three well-separated identical peaks (BEC
droplets), each one being practically indistinguishable from the
above-mentioned fundamental GS. Figure \ref{Fig4}(c) shows that the results
obtained from the effective 1D equation (\ref{II-18}) (the solid red line)
agree very well with those produced by the full 3D GPE (\ref{II-17}) (open
circles). In particular, the 1D equation yields the particle number $N=56$,
which is very close to $N=55$, as given by the 3D solution. Since $N$
represents a measurement of the norm of the wave function, it is clear that
the error in the number of particles reflects the error in the corresponding
wave functions. The 1D GPE (\ref{II-18b}), corresponding to the dashed blue
line in Fig. \ref{Fig4}(c), gives $N=49$, which implies an error $\sim10\%$.


\begin{figure}[ptb]
\begin{center}
\includegraphics[
width=8.2cm
]{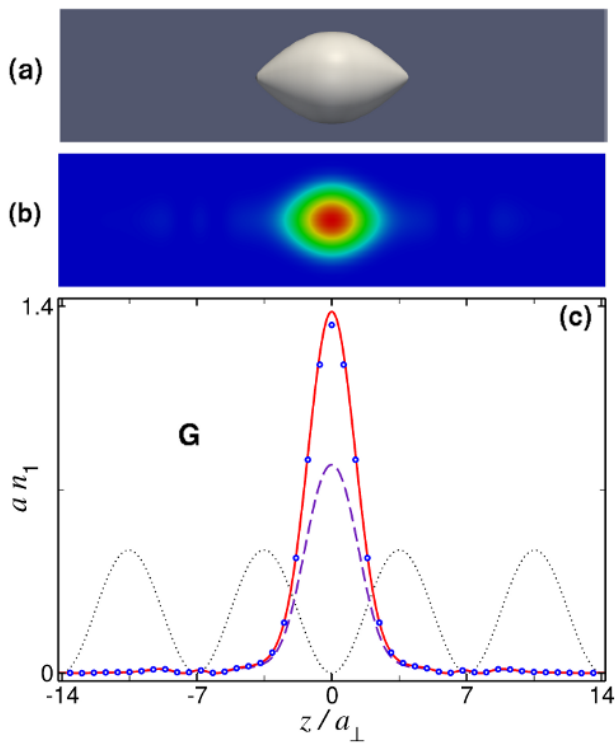}
\end{center}
\caption{(Color online) Same as Fig. \ref{Fig2} but for the gap soliton
corresponding to point G in Fig. \ref{Fig1}(c).}%
\label{Fig5}%
\end{figure}


Point F in Fig. \ref{Fig1}(c) corresponds to a fundamental GS located near the
top edge of the first band gap. This soliton, which is very similar to that
shown in Fig. \ref{Fig5}, contains $71$ particles and has chemical potential
$\widetilde{\mu}=11.5$, which implies $\mu=2.15\,\hbar\omega_{\bot}$. At this
value of $\mu$ there is a significant probability of the excitation of higher
levels in the radial-confinement potential, which is corroborated by the fact
that the peak axial density of this GS is $an_{1}(0)\simeq0.7$. Because
inequality $an_{1}\ll1$ does not hold in this case, one cannot expect the
ordinary 1D GPE (\ref{II-18b}) to be valid. In fact, it yields the number of
particles $N=53$, which implies an error greater than $25\%$. Nevertheless,
the effective nonpolynomial 1D equation (\ref{II-18}) remains
valid and quite accurate. It produces $N=73$, which corresponds to a
maximum error of $2.8\%$. This is not surprising, since condition
(\ref{II-14}), which guarantees the validity of the nonpolynomial equation, is
satisfied in this case. \newline

The solution pertaining to point G in Fig. \ref{Fig1}(c) is qualitatively
similar. It corresponds to the fundamental GS located near the top edge of the
second band gap, with chemical potential $\widetilde{\mu}=17.2$ and $N=182$.
Figure \ref{Fig5} shows the 3D density and axial linear density $an_{1}(z)$ of
this BEC droplet. In this case, $an_{1}(0)\simeq1.3$, indicating a large
contribution of excited radial modes. As a consequence, the ordinary 1D GPE
(\ref{II-18b}) [the dashed blue line in Fig. \ref{Fig5}(c)] fails to reproduce
the 3D results (open circles). It predicts $N=119$, which means the error exceeds 
$34\%$. As seen from Fig. \ref{Fig5}(c), the effective 1D equation
(\ref{II-18}) (the solid red line) remains accurate enough in this case, too.
In particular, it yields $N=187$, the respective error being $2.8\%$.

The above results imply that, with the deep OL, the number of particles that
can be accommodated in a fundamental GS can be large enough to make the 3D
character of the wave function essential. In fact, even for the tight radial
confinement ($E_{R}/\hbar\omega_{\bot}\ll1$), which is the most favorable case
for the applicability of the 1D approximation, the usual 1D GPE (\ref{II-18b})
with the deep OL potential cannot be reliably used beyond the first third of
the first band gap. Since the validity of the mean field treatment requires
$N\gg1$, the usual GPE cannot be used close to the bottom edge of the band gap, 
either. It is thus clear that the range of validity of this standard 1D
equation is limited. This conclusion becomes even more severe if one takes
into account that the tight-radial-confinement regime is not easy to realize
using typical experimental parameters. Our simulations indicate, however, that
the effective nonpolynomial 1D equation (\ref{II-18}) properly describes such
essentially 3D situations, providing for an accurate description of the
fundamental GSs in the entire band gaps.


\begin{figure}[ptb]
\begin{center}
\includegraphics[
width=8.2cm
]{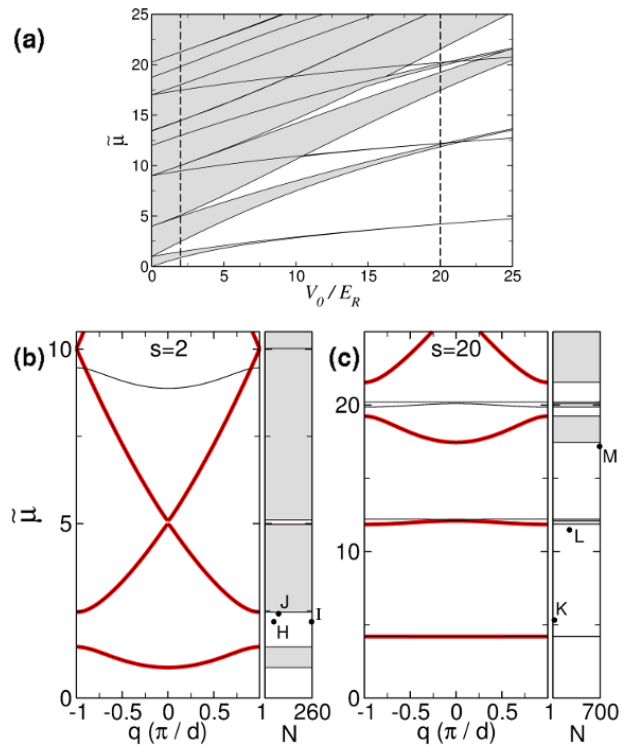}
\end{center}
\caption{(Color online) (a) Band-gap structure of a noninteracting 3D BEC
with $E_{R}/\hbar\omega_{\bot}=1/4$, as a function of the dimensionless
lattice depth $s\equiv V_{0}/E_{R}$. The representative cases $s=2$ and $20$,
indicated by vertical dashed lines, are considered in more detail in (b) and
(c), respectively, which show the dimensionless chemical potential
$\widetilde{\mu}$ as a function of the quasimomentum $q$ in the first
Brillouin zone (in units of $\pi/d$). The right panels display the location of
the $^{87}$Rb gap solitons considered in this work (points H--M).}%
\label{Fig6}%
\end{figure}


\subsection{Intermediate radial confinement: $E_{R}/\hbar\omega_{\bot
}\simeq1/4$}

This regime is of particular interest because it corresponds to typical
experimental parameters. It can be realized, for instance, with a $^{87}$Rb
condensate in an OL of period $d=1.55$ $\operatorname{\mu m}$, radially
confined by a harmonic potential with $\omega_{\bot}/2\pi=960$
$\operatorname{Hz}$. The particle numbers $N$ of the GSs considered below are
given for these physical parameters, and, they can be converted into values of
$N$ for other situations by means of Eq. (\ref{II-19c}) (now, with $a/a_{\bot
}=0.015$) .

Figure \ref{Fig6}(a) shows the band-gap structure produced by the
linearized version of Eq. (\ref{II-17}) for zero-vorticity modes. Recall that
the linearization of the effective 1D equations (\ref{II-18}) and
(\ref{II-18b}) yields, instead, the diagram displayed in Fig. \ref{Fig1}(a)
(except for the narrow band marked by the arrow, which, as already mentioned, 
does not appear in the 1D approximation). It is seen that the 1D approximation
cannot reproduce the 3D picture resulting from the excitation of the radial
modes, which manifest themselves in Fig. \ref{Fig6}(a) as replicas of bands
shifted up by integer multiples of $2\hbar\omega_{\bot}/E_{R}$. Since\ this
quantity is smaller in the present case than it was before, the effect of
these contributions in the region of interest is now stronger. While it is
obvious that the 1D equations cannot account for the 3D effects originating
from the shifted bands, we will demonstrate that these equations may still
produce an accurate description of common GSs (i.e., those originating from
the $m=n_{r}=0$ energy bands).

Figures \ref{Fig6}(b) and \ref{Fig6}(c) display the band-gap structure as a 
function of the quasimomentum for lattice depths $V_{0}/E_{R}=2$ and $20$, 
respectively.
Thin black lines in these figures represent 3D results that cannot be
reproduced by the 1D models, and points H--M in $\left(  N,\widetilde{\mu
}\right)  $ plane mark examples of GSs that will be considered below.

\subsubsection{Shallow optical lattice: $V_{0}/E_{R}=2$}

\begin{figure}[ptb]
\begin{center}
\includegraphics[
width=8.2cm
]{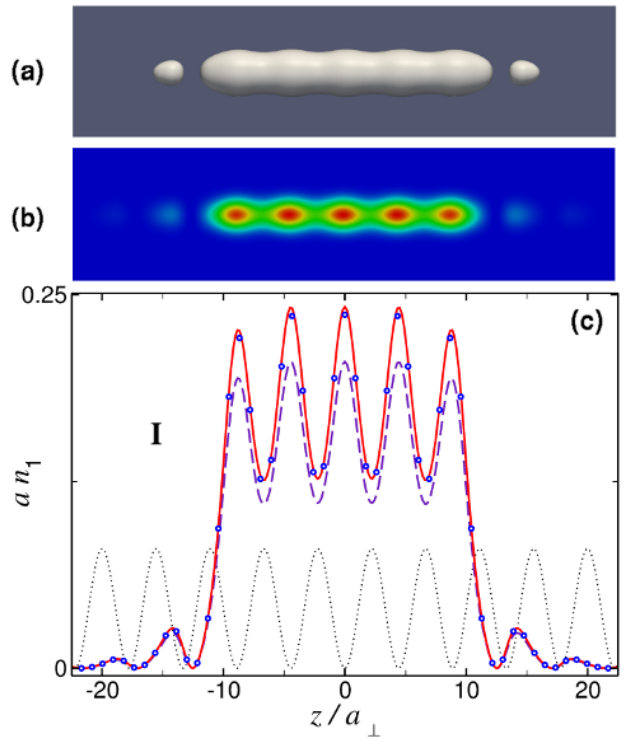}
\end{center}
\caption{(Color online) Atom density of the gap soliton corresponding to
point I in Fig. \ref{Fig6}(b), displayed as (a) an isosurface taken at $5\%$
of the maximum density and (b) as a color map along a cutting plane
containing the $z$ axis. (c) Dimensionless axial density $an_{1}$ obtained
from the 3D wave function, as prescribed by Eq. (\ref{II-2}) (open circles)
along with the corresponding predictions from the nonpolynomial 1D equation
(\ref{II-18}) (solid red line) and the ordinary 1D GPE (\ref{II-18b}) (dashed
blue line).}%
\label{Fig7}%
\end{figure}

Point H in Fig. \ref{Fig6}(b) corresponds to a fundamental GS with
$\widetilde{\mu}=2.19$ and $N=50$, which is similar to the one in Fig.
\ref{Fig3}, but with the peak axial density $an_{1}(0)\simeq0.2$. In this
case, the effective 1D equation (\ref{II-18}) predicts $N=51$, thus
corresponding to a $2\%$ error, while the 1D GPE (\ref{II-18b}) predicts
$N=45$ (an $11\%$ error). Symmetric combinations of five such fundamental GSs
generate the compound GS shown in Fig. \ref{Fig7}, which contains $258$
particles and corresponds to point I in Fig. \ref{Fig6}(b). The 3D density of
this soliton exhibits five weakly separated peaks, as shown in Figs.
\ref{Fig7}(a) and \ref{Fig7}(b) by means of the isosurface and the color map in 
the cross-section plane drawn through the $z$ axis. As seen in Fig. \ref{Fig7}(c),
in this case the effective 1D equation (\ref{II-18}) yields an error of
$1.5\%$ in the norm of the wave function ($N=262$) while the use the ordinary
1D GPE (\ref{II-18b}) generates an error of $12\%$ ($N=226$), showing that,
already for this GS, the 3D effects play an important role.

Point J in Fig. \ref{Fig6}(b) corresponds to a fundamental GS containing $75$
particles and located near the top edge of the first band gap. It is
qualitatively similar to the one shown in Fig. \ref{Fig3}, but having
$\widetilde{\mu}=2.42$ and $an_{1}(0)\simeq0.26$. In this case, Eqs.
(\ref{II-18}) and (\ref{II-18b}) yield results with an error $\simeq1\%$
($N=76$) and $12\%$ ($N=66$), respectively.


\subsubsection{Deep optical lattice: $V_{0}/E_{R}=20$}

Points K, L, and M in Fig. \ref{Fig6}(c) correspond to fundamental GSs in the
case of the tight-binding underlying linear structure. The first soliton,
which contains $19$ particles, with $\widetilde{\mu}=5.33$ and axial density
$an_{1}(0)\simeq0.23$, also corresponds to the nonlinear tight-binding regime
and is thus highly localized at a single lattice site. This follows from the
fact that, for $V_{0}/E_{R}=20$ and $E_{R}/\hbar\omega_{\bot}=1/4 $, one has
$V_{0}/\,\hbar\omega_{\bot}=5$, which is much greater than the above-mentioned
value of $an_{1}(0)$. For this GS, located near the bottom edge of the first
band gap, the ordinary 1D GPE (\ref{II-18b}) yields an error $>11\%$, which
continues to grow as one moves upward in the band gap. However, the effective
nonpolynomial 1D equation (\ref{II-18}) remains accurate within a $2.5\%$ deviation.


\begin{figure}[ptb]
\begin{center}
\includegraphics[
width=8.2cm
]{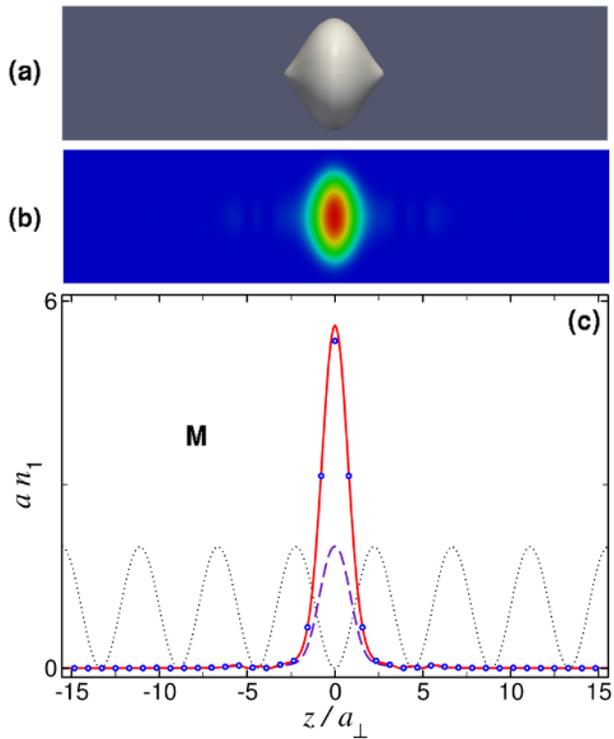}
\end{center}
\caption{(Color online) Same as Fig. \ref{Fig7} but for the gap soliton
corresponding to point M in Fig. \ref{Fig6}(c).}%
\label{Fig8}%
\end{figure}


Point L in Fig. \ref{Fig6}(c) indicates the position of a fundamental GS near
the top edge of the first band gap, with $\widetilde{\mu}=11.5$, $N=243$, and
$an_{1}(0)\simeq2.3$. Since the inequality $an_{1}\ll1$\ does not hold in this
case, the ordinary 1D GPE (\ref{II-18b}) is invalid, giving a $45\%$ error
($N=133$). Note that, even though point L is relatively close to a 3D energy
band [the thin black line in Fig. \ref{Fig6}(c)], which cannot be reproduced
by the effective 1D equation (\ref{II-18}), it can, however, reproduce this GS
within a $4\%$ error ($N=253$). The same is true for the fundamental GS
displayed in Fig. \ref{Fig8}, which corresponds to point M in Fig.
\ref{Fig6}(c). It represents a BEC droplet containing $696$ particles, with
$\widetilde{\mu}=17.2$ and the peak axial density $an_{1}(0)\simeq5.4$. Since
we have $V_{0}/\,\hbar\omega_{\bot}<an_{1}(0)$ in the present case, the system
is no longer in the nonlinear tight-binding regime. This is seen in Fig.
\ref{Fig8}, where the density is not strongly localized around the minimum of
the potential well. As seen in Fig. \ref{Fig8}(c), the nonpolynomial 1D
equation (\ref{II-18}) yields results (the solid red line) that agree with the
3D picture produced by the full GPE (\ref{II-17}) (open circles), within a
$3\%$ deviation [Eq. (\ref{II-18}) yields $N=719$ in this case]. On the
contrary, the 1D GPE (\ref{II-18b}), which gives results (the dashed blue
line) with an error $>55\%$ (it yields $N=297$), clearly is not valid in this case.

These findings demonstrate that, in the physically relevant regime of the
intermediate radial confinement ($E_{R}/\hbar\omega_{\bot}\gtrsim1/4$), even
for the shallow OL the 3D effects may be important, and thus the usual 1D GPE
(\ref{II-18b}) fails to reproduce correctly the axial density $an_{1}(z)$ and
the particle content $N$. For $E_{R}/\hbar\omega_{\bot}=1/4$ and potential
depth $V_{0}/E_{R}=2$, the fundamental GSs located in the first band gap, as
predicted by the 1D GPE equation, feature the error $\simeq12\%$, which is
still larger for compound GSs. For the deep OL ($V_{0}/E_{R}=20$), the 1D GPE
is valid only in a small region near the bottom edge of the first band gap. In
general, this equation is applicable only where both conditions $an_{1}\ll1$
and $N\gg1$ hold simultaneously. As $E_{R}/\hbar\omega_{\bot}$ increases, the
relative strength of the radial confinement decreases, and the range of
validity of the 1D GPE becomes more and more narrow. From the experimental
viewpoint, the increase of $E_{R}/\hbar\omega_{\bot}$ may be relevant because, 
in this way, one can easily increase the number of particles in the fundamental
GSs. However, the increase of $E_{R}/\hbar\omega_{\bot}$ also implies a
decrease in the relative size of the radial-excitation energy quantum, which
can compromise the stability of the solitons because they can decay by
exciting higher radial modes, even for condensates with a relatively small
number of particles. We briefly consider this case below. The stability
properties of the GSs will be analyzed in Sec. IV.


\begin{figure}[ptb]
\begin{center}
\includegraphics[
width=8.2cm
]{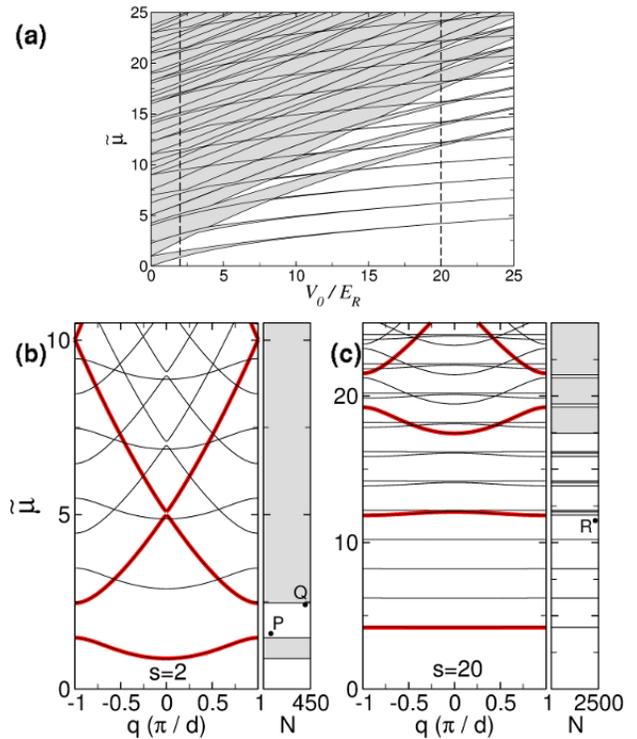}
\end{center}
\caption{(Color online) (a) Band-gap structure of a noninteracting 3D BEC
with $E_{R}/\hbar\omega_{\bot}=1$, as a function of the dimensionless lattice
depth $s\equiv V_{0}/E_{R}$. The representative cases $s=2$ and $20$,
indicated by vertical dashed lines, are considered in more detail in (b) and
(c), respectively, which show the dimensionless chemical potential
$\widetilde{\mu}$ as a function of the quasimomentum $q$ in the first
Brillouin zone (in units of $\pi/d$). The right panels display the location of
the $^{87}$Rb gap solitons considered in this work (points P--R).}%
\label{Fig9}%
\end{figure}


\subsection{Weak radial confinement: $E_{R}/\hbar\omega_{\bot}\geq1$}

In this regime, the typical energy scale in the underlying linear problem is
sufficiently large to easily excite higher transverse modes. As a
representative example, we consider the case of $E_{R}/\hbar\omega_{\bot}=1$,
which can be realized in a $^{87}$Rb condensate in an OL with $d=1.55$
$\operatorname{\mu m}$ and radial-confinement strength $\omega_{\bot}%
/2\pi=240$ $\operatorname{Hz}$. The particle contents of the GSs considered
below correspond to such condensates. As before, the corresponding values of
$N$ can be converted into those corresponding to other situations by means of
Eq. (\ref{II-19c}) (this time, $a/a_{\bot}=7.6\times10^{-3}$).

Figure \ref{Fig9}(a) displays the band-gap structure of the underlying 3D
linear problem, to be compared with Fig. \ref{Fig1}(a) which shows the band-gap
structure obtained from the linearization of 1D equations (\ref{II-18}) or
(\ref{II-18b}). As before, the latter equations cannot account for the 3D
contributions from the excited radial modes, which account for the series of
shifted bands separated by gaps of value $2\hbar\omega_{\bot}/E_{R}$ in Fig.
\ref{Fig9}(a). Figures \ref{Fig9}(b) and \ref{Fig9}(c) display the linear band-gap
structure as a function of the quasimomentum for $V_{0}/E_{R}=2$ and $20$, respectively.


\begin{figure}[ptb]
\begin{center}
\includegraphics[
width=8.2cm
]{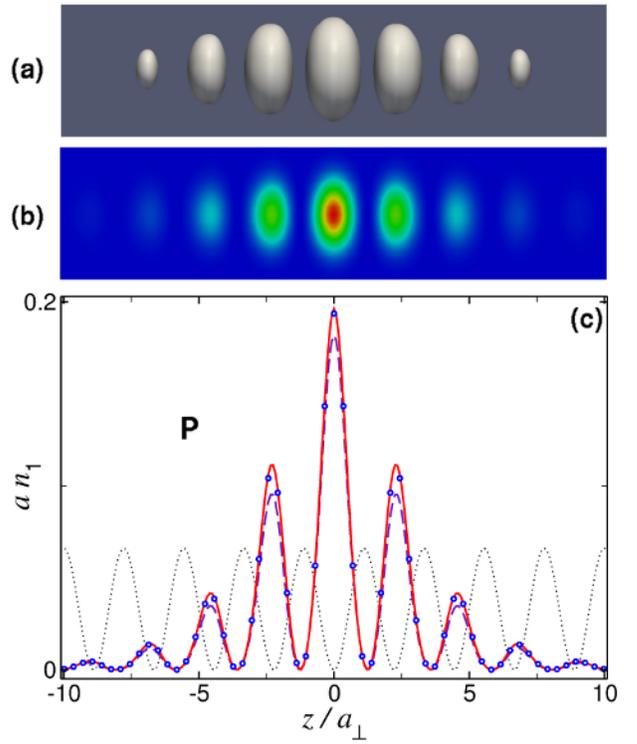}
\end{center}
\caption{(Color online) Atom density of the gap soliton corresponding to
point P in Fig. \ref{Fig9}(b), displayed as (a) an isosurface taken at $5\%$
of the maximum density and (b) as a color map along a cutting plane
containing the $z$ axis. (c) Dimensionless axial density $an_{1}$ obtained
from the 3D wave function, as prescribed by Eq. (\ref{II-2}) (open circles)
along with the corresponding predictions from the nonpolynomial 1D equation
(\ref{II-18}) (solid red line) and the ordinary 1D GPE (\ref{II-18b}) (dashed
blue line).}%
\label{Fig10}%
\end{figure}


Point P in Fig. \ref{Fig9}(b) corresponds to a fundamental GS with
$\widetilde{\mu}=1.59$ and $N=74$. Its 3D density plot, shown in Fig.
\ref{Fig10}, demonstrates that this fundamental soliton is spread over several
lattice sites, which is a consequence of the fact that its chemical potential
is very close to the first linear energy band, where only extended solutions
of the stationary GPE exist. Figure \ref{Fig10}(c) shows that the results
obtained from the effective 1D equation (\ref{II-18}) (the solid red line)
coincide with the 3D results (open circles) within $1\%$ (it predicts $N=75$).
However, the 1D GPE (\ref{II-18b}) (the dashed blue line) gives rise to an
error $\simeq10\%$ (it predicts $N=66$). We thus conclude that, in the
weak-radial-confinement regime ($E_{R}/\hbar\omega_{\bot}\geq1$), the latter
equation is only applicable in a narrow region close to the bottom edge of the
first band gap, even for shallow OLs. In this regime, the 3D contributions play
an important role in most cases. These contributions are, however, well
accounted for by the effective 1D equation (\ref{II-18}). For instance, for
point Q in Fig. \ref{Fig9}(b) [which corresponds to a fundamental GS near the
top edge of the first band gap, with $\widetilde{\mu}=2.42$, $N=400$, and
$an_{1}(0)\simeq1.5$] the 1D GPE (\ref{II-18b}) gives an error of $34\%$
($N=265$), while the nonpolynomial 1D equation (\ref{II-18}) limits the error
to $3.5\%$ ($N=414$). Point R in Fig. \ref{Fig9}(c) is an example of a
fundamental GS in a deep OL. It corresponds to a disk-shaped BEC droplet
trapped in a single lattice cell [see Fig. \ref{Fig11}(b)], which contains
$2323$ $^{87}$Rb atoms, and has $\widetilde{\mu}=11.5$. Its axial linear
density is qualitatively similar to that shown in Fig. \ref{Fig8}(c), but with
a maximum value $an_{1}(0)\simeq23$. In this case, the 1D GPE predicts the
particle content $N=534$, while the effective 1D equation (\ref{II-18}) yields
$N=2437$, which corresponds to an error $\simeq5\%$ in the norm of the wave
function. Thus, the 1D nonpolynomial equation provides for a sufficiently good
description of the stationary GSs even in the highly nonlinear ($an_{1}\gg1$)
weak-radial-confinement regime.

\section{STABILITY ANALYSIS}

We have investigated the stability of the GS solutions by monitoring their
long-time behavior after the application of a random perturbation
\cite{Wu2,JYang}. Specifically, we have perturbed the corresponding wave
functions with a small-amplitude ($\sim2\%$) additive Gaussian white noise and
have monitored the subsequent nonlinear evolution for $1$ $\operatorname{s}$.
To this end we have numerically solved the full 3D GPE (\ref{II-0a}), as well
as the effective 1D equation (\ref{II-13}), using a Laguerre-Fourier
pseudospectral method with a third-order Adams-Bashforth time-marching scheme.
Numerical integration of the 3D GPE for such a long time is a very demanding
computational task. While it is possible for a certain GS to be metastable and
decay on a time scale still longer than $1$ $\operatorname{s}$, in practice
this time is long enough in comparison with the lifetime of the condensate per
se, hence the soliton surviving for $1$ $\operatorname{s}$ may be categorized
as stable.


\begin{figure}[ptb]
\begin{center}
\includegraphics[
width=8.2cm
]{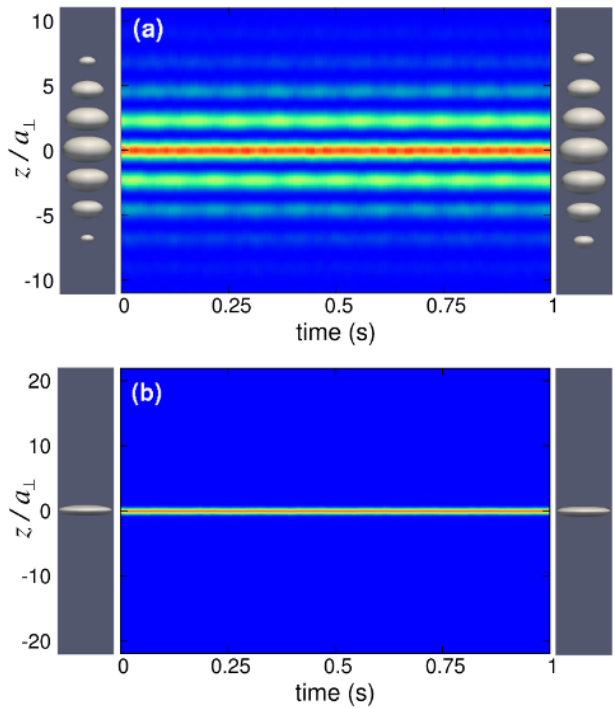}
\end{center}
\caption{(Color online) Evolution in time, after the application of a $2\%$
white-noise perturbation, of the fundamental gap solitons corresponding (a) to
point P in Fig. \ref{Fig9}(b) and (b) to point R in Fig. \ref{Fig9}(c).}%
\label{Fig11}%
\end{figure}


We have found that both the full 3D GPE (\ref{II-0a}) and the effective
nonpolynomial 1D equation (\ref{II-13}) lead to the same conclusions regarding
the stability of the GSs. However, once an unstable soliton begins to decay,
one cannot expect, in general, the latter equation to reproduce the dynamical
behavior correctly. The reason for this problem is that, as already mentioned 
in Sec. II, in the presence of the OL potential the above-mentioned adiabatic
condition (i) is hard to fulfill in time-dependent settings. Of course, the
same is true for the 1D equation (\ref{II-0c}), which is a limit form of Eq.
(\ref{II-13}) and, consequently, has a much narrower range of applicability.
The results presented below have been obtained using the full 3D GPE
(\ref{II-0a}). For each GS we have used at least two different basis sets and
time steps to check that the results do not depend on details of the numerical procedure.

Our simulations demonstrate that, in general, $(1,0,0)$ fundamental GSs are
stable, except in a narrow region close to the top edge of the band gaps. In
particular, the solitons corresponding to points A, D, K, and P in Figs.
\ref{Fig1}, \ref{Fig6}, and \ref{Fig9} remain stable up to $t=1$
$\operatorname{s}$. On the contrary, GSs in a shallow OL ($V_{0}/E_{R}=2$),
located close to the top edge of a band gap, such as those corresponding to
points C, H, J, and Q in Figs. \ref{Fig1}, \ref{Fig6}, and \ref{Fig9}, turn
out to be unstable. This instability manifests itself as a steady decay of the
norm of the GS through emission of radiation, on a time scale that increases
as one moves away from the top edge of the band gap. In this regard, one finds
that the GS corresponding to point H decays much slower than the other ones.
As the lattice depth $V_{0}/E_{R}$ increases, in general, GSs become more
stable and the instability region shrinks. In particular, our simulations
indicate that GSs such as those corresponding to points F, G, L, M and R in
Figs. \ref{Fig1}, \ref{Fig6}, and \ref{Fig9} (which are fundamental GSs in the
deep OL, with $V_{0}/E_{R}=20$, located close to the top edge of the band gap)
also remain stable up to $t=1$ $\operatorname{s}$. This is in good agreement
with previous analyses carried out in the context of deep optical lattices in
terms of the ordinary 1D GPE (\ref{II-0c}) \cite{Kiv1,Wu2}.


\begin{figure}[ptb]
\begin{center}
\includegraphics[
width=8.2cm
]{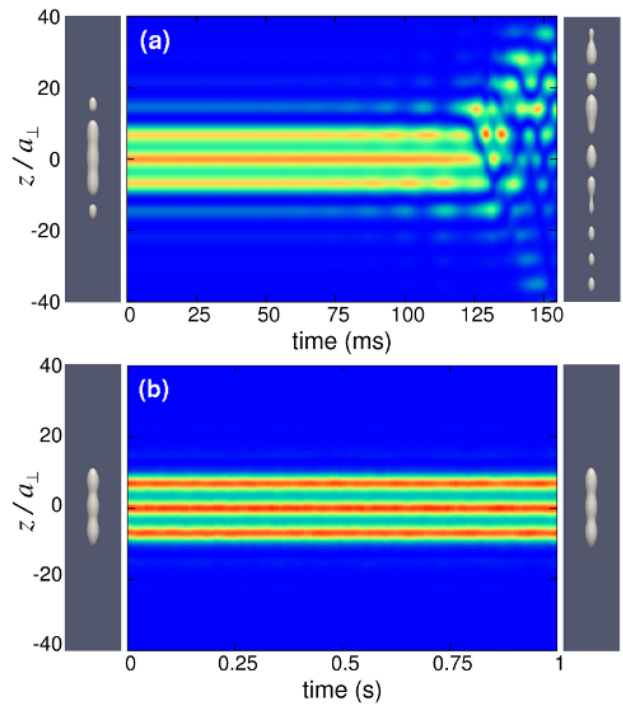}
\end{center}
\caption{(Color online) (a) Evolution in time, after the application of a
$2\%$ white-noise perturbation, of the compound gap soliton corresponding to
point B in Fig. \ref{Fig1}(b). Panel (b) displays the long-time behavior,
after the application of the perturbation, of a stable three-peak soliton (see
text).}%
\label{Fig12}%
\end{figure}


An important result is that $(1,0,0)$ fundamental GSs remain stable even in
the weak-radial-confinement regime ($E_{R}/\hbar\omega_{\bot}\simeq1$). In
this regime, GSs always have sufficient energy to excite higher radial modes,
and, as a consequence, the 3D effects are always relevant and the usual 1D GPE
(\ref{II-0c}) fails. Figure \ref{Fig11}(a) displays the long-time behavior,
after the application of the perturbation, of the fundamental GS shown in Fig.
\ref{Fig10} [it corresponds to point P in Fig. \ref{Fig9}(b)]. This figure
shows (by means of a color map) the evolution of the axial density
$an_{1}(z,t)$, which has been obtained from the 3D wave function
$\psi(\mathbf{r}_{\bot},z,t)$ by integrating out the radial dependence, as per
Eq. (\ref{II-2}). The left panel in this figure shows the 3D condensate
density at $t=0$ (i.e., just after the application of the perturbation) as an
isosurface taken at $5\%$ of the maximum density, while the right panel
represents the density at $t=1$ $\operatorname{s}$. This GS has $\widetilde
{\mu}=1.59$, which implies $\mu=2.59\hbar\omega_{\bot}$. Note that, despite
the fact that this chemical potential is greater than the quantum $\hbar
\omega_{\bot}$, the GS remains stable up to $1$ $\operatorname{s}$. The same
is true for the fundamental GS corresponding to point R in Fig. \ref{Fig9}(c),
which corresponds to a disk-shaped BEC containing more than $2300$ $^{87}$Rb
atoms with chemical potential $\mu=12.5\hbar\omega_{\bot}\gg\hbar\omega_{\bot
}$. As Fig. \ref{Fig11}(b) shows, this GS also remains stable.

Our simulations also demonstrate that the compound GS from Fig. \ref{Fig2},
corresponding to point B in Fig. \ref{Fig1}(b), is unstable. This is seen in
Fig. \ref{Fig12}(a), which shows how the soliton decays on a time scale
$\simeq120$ $\operatorname{ms}$ after the addition of a $2\%$ white-noise
perturbation. This compound GS, which has $\widetilde{\mu}=1.75$ and $N=35$,
was built in the shallow OL ($V_{0}/E_{R}=2$) as the symmetric combination of
three fundamental constituents. We have found that, in such shallow lattices,
GSs of this type are always unstable. However, it is not difficult to find
stable solitons of this type in somewhat deeper lattices. An example is shown
in Fig. \ref{Fig12}(b), which represents a stable three-peak soliton with
$\widetilde{\mu}=3.13$ and $N=59$, trapped in the OL with $V_{0}/E_{R}=4$ and
$E_{R}/\hbar\omega_{\bot}=1/10$. Similar dynamics is observed for the GS
displayed in Fig. \ref{Fig7}, which corresponds to point I in Fig.
\ref{Fig6}(b). This is a five-peak compound generated in the shallow OL
($V_{0}/E_{R}=2$) in the regime of the intermediate radial-confinement
strength ($E_{R}/\hbar\omega_{\bot}\simeq1/4$). Our simulations indicate that
this compound soliton is unstable. In fact, no stable five-peak solitons were
found in such a shallow lattice. On the other hand, it is not difficult to
find stable compounds of this kind for deeper OLs. An example is the five-peak
pattern with $V_{0}/E_{R}=4$, $E_{R}/\hbar\omega_{\bot}=1/4$, $\widetilde
{\mu}=2.83$, and $N=212$. In general, GSs naturally become more stable as the
lattice depth increases. For deep OLs ($V_{0}/E_{R}=20$), the stability of
compound GSs is essentially identical to that of their fundamental
constituents. The GS in Fig. \ref{Fig4}, which corresponds to point E in Fig.
\ref{Fig1}(c), is an example of a stable three-peak soliton realized in a deep OL.

\section{CONCLUSIONS}

To appraise the physical relevance of the results reported in this work, it is
pertinent to recall that the mean-field treatment of the GSs (gap solitons),
based on the GPE, is valid if $N\gg1$ and the condensate is sufficiently
dilute so that $a^{3}\overline{n}\ll1$, where $a$ is the scattering length of
the inter-atomic collisions and $\overline{n}$ is the atomic density. In
shallow OLs (optical lattices) such conditions can be easily met, though the
former one imposes a serious limitation on the range of applicability of the
usual 1D GPE (\ref{II-18b}), which fails as $N$ increases. In deep OLs, where
the tunneling between adjacent cells is strongly suppressed, the above
conditions must be fulfilled at each site of the lattice. Under these
circumstances, GSs may have a rather high local density. A good estimate for
the 3D peak density $n(\mathbf{0})=N|\psi(\mathbf{0})|^{2}$ in terms of the
peak axial density, $n_{1}(0)$, can be obtained from the following relation
\cite{PRA77}:%
\begin{equation}
n(\mathbf{0})=\frac{n_{1}(0)}{\pi a_{\bot}^{2}\sqrt{1+4an_{1}(0)}}.
\label{IV-1}%
\end{equation}
Substituting the values of $an_{1}(0)$ obtained in Section III for different
GSs into Eq. (\ref{IV-1}), one can easily verify that condition $a^{3}%
n(\mathbf{0})\ll1$ is satisfied in all cases, which justifies the use of the
mean-field description. The most critical situation occurs for the GS
corresponding to point G in Fig. \ref{Fig1}(c), which has $a^{3}%
n(\mathbf{0})\simeq10^{-4}$. For the GS corresponding to point R in Fig.
\ref{Fig9}(c), one finds $a^{3}n(\mathbf{0})\simeq5\times10^{-5}$.

Despite the fact that applicability conditions for the mean-field treatment
can be readily met, it is much harder to justify the validity of the usual 1D
GPE (\ref{II-0c}), which requires both conditions $an_{1}\ll1$ and
$N\gg1$ to hold simultaneously. Only in this case may the 3D effects be
safety neglected. In most experimentally relevant situations, these
conditions\ are not met, hence the applicability range of Eq. (\ref{II-0c})
turns out to be very limited. On the contrary, it has been shown in this work
that the effective 1D GPE (\ref{II-13}) with the nonpolynomial nonlinearity
provides for an accurate description of the stationary matter-wave GSs in most
cases of practical interest.

A relevant extension of the present analysis may be to GSs with intrinsic
vorticity, as well as to mobility of stable solitons. It is also interesting
to consider two-component GSs in the realistic 3D setting.

\begin{acknowledgments}
V.D. acknowledges financial support from Ministerio de Ciencia e
Innovaci\'{o}n through contract No. FIS2009-07890 (Spain).
\end{acknowledgments}

\end{document}